# Designing Survival Strategies for Propulsion Innovations

**Thomas Aigle**
Fuel Cell Education and Training Center Ulm (WBZU)
Helmholtzstraße 6
89081 Ulm
Germany
Email: <thomas.aigle@wbzu.de>

**Lutz Marz**
Social Science Research Center Berlin (WBZ)
Reichpietschufer 50
10785 Berlin
Germany
Email: <lutz@wzb.eu>

**Andrea Scharnhorst (corresponding author)**
The Virtual Knowledge Studio for the Humanities and Social Sciences (VKS)
Royal Netherlands Academy of Arts and Sciences (KNAW)
Cruquiusweg 31
1019 AT Amsterdam
The Netherlands
Email: <andrea.scharnhorst@vks.knaw.nl>

# Designing Survival Strategies for Propulsion Innovations

## Abstract

Mobility is valued greatly in the highly industrialized societies. The need for radical change in propulsion technologies is obvious to all actors, irrespective of whether they originate from industry, politics or the general public. This paper analyses the tension between innovation pressure and pull of convention in the automobile industries. This tension is currently giving rise to a situation of stalemate in relation to alternative propulsion and fuel technologies. We map the situation by means of a taxonomy of current and future incremental and radical innovations. Based on in-depth field observation of engineering and manufacturing in Germany, we present an innovation landscape in the form of a two-dimensional matrix composed of propulsion innovations and fuel innovations. We use mathematical models of hyperselection to develop a rationale for escape strategies from the current lock-in into conventional combustion-engine technology. Based on the heuristic guidance of these models, we discuss several empirical cases in which buses act as pioneers in markets for alternative propulsion vehicles. Neither the model nor the basic empirical material used in this paper are new. Instead, we show that using mathematical models as a kind of substratum on which empirical observations and theoretical arguments can be ordered leads to new and partly unexpected insights into the nature of socioeconomic innovations. We apply this approach to the case of new driving technologies and argue that there is a "third epistemic function" of models in the continuum between data driven, exactly validated models at the one extreme and a metaphoric use of models in thought experiments at the other. This is what we call a holistic description.

# 1. Introduction: Paradigm change, the pull of convention and innovation pressure

## 1.1 Paradigm change

A paradigm change away from fossil fuel technologies to regenerative energy technologies will take place throughout the world in the 21st century [1, 2]. This paradigm change concerns the very foundations of modern societies. This is demonstrated very clearly in one of the main sectors of these societies, i.e. that of automobility, which includes all road-related passenger and goods transport (passenger cars, buses, heavy goods vehicles etc.). Within this sector, all propulsion and fuel systems, including the associated production and service infrastructures, will have to be converted from fossil to regenerative energy sources [3].

The energy-technology paradigm change in the area of automobility constitutes a problem of existential import for modern societies. The automobile is by far the preferred mode of transport over distances: approximately 90 percent of all passenger kilometres that can be covered in a mode of transport are accounted for by the car [4 (p. 9)]. However, that is not the full story. The car is far more than a simple means of conveyance. It is an integral component of the fossil-fuel-based mobility society and constitutes a symbol of status, prosperity, freedom (of movement) and the dynamics and progress of the modern age [5]. This is precisely how the car is perceived in the emerging growth societies, for example China, where it is celebrated as both an individual status symbol and proof of the modernity of society [6].

The question of the future of individual mobility arises anew against the background of a paradigm change in energy policy. As recently as 1996, Cowan and Hulten [7] stated in their study on the future of the electric car that:

> It seems clear that a rapid escape from lock-in, a move from gasoline to the electric vehicle, is not going to happen. In the present climate, the electric vehicle has to compete with the gas vehicle under conditions established by users' 90-year relationship with the gas car, and there are technical problems that make the electric vehicle inferior (p. 77).

In this paper we re-address the question as to how the current lock-in in the automobile industries can be escaped by examining both the conventional and alternative technological pathways.

The challenges that arise from the paradigm change in energy technology for both the automobile industry, in particular, and society, in general, have been examined from different conceptual perspectives in a wide range of studies, for example in relation to globalisation effects and the dynamic effects in the car industry [8] and in relation to the acceptance [9, 10] and use [11] of alternative means of propulsion. Several studies examine specific alternative propulsion technologies such as hydrogen cars [12]. The relationship between mobility and energy markets (so-called vehicle-to-grid systems) is also considered in a recent study about the possible implications for the survival of electric vehicles [13]. Based on a scenario framework coupling energy, climate, user behaviour and technological change, a structural substitution of conventional internal combustion engine vehicles is forecast to take place not earlier than 2030 [14]. Another study has mapped technological trajectories of fuel cell research in the patent space [64]. In our study we design a space of technological innovation for vehicles based on qualitative research that is as comprehensive as possible. Both existing and future technologies can be located in this space along the engine and fuel axes. Starting from such a technological space we describe and visualize the dominant designs in the automobile industry. We use dynamic models to find survival strategies for early adopters who aim to break out of the deadlock.

Like Cowan and Hulten [7], in their study on electric vehicles, we start from the well-known fact that the automobile industry is currently in a classical lock-in situation in terms of both technology [15] and the market [16]. As compared with other branches of industry, the automobile industry has been characterized as very important on the basis of its market size and complexity [17, 18]. There are many good reasons why innovation studies have focussed on the automobile industry almost since its inception: to name but a few, it is one of major dominant manufacturing industries which served as prototype for many other manufacturing industries over the past century; it is old enough to allow analysis of the type of innovation that dominates at different stages in the life of an industry; and it is an excellent example of networked production based on a network of suppliers [19, 20, 21, 64]. The presentation of a comprehensive review of the literature in this area would far exceed the aim of this study. Thus, we concentrate, in particular, on literature that is relevant to the mapping of the innovation space and design of future scenarios.

Frenken et al. [22] examined the situation of low emission vehicles several years ago and noted:

> The current state of development in environmentally friendly automotive propulsion is best characterized as uncertain: the development potential of various options is high, yet it is unclear which option is optimal from an environmental and economic point of view (p. 493).

This statement remains valid [23]. There are two reasons for the continuing high level of uncertainty which appears to be blocking strategic decision making:

- the enormous variety of alternative technologies, which makes it extremely difficult to predict the next winning technology, and
- the dominance of the "old" conventional technology.

Frenken et al.'s [22] analysis focussed on a possible premature lock-in into a suboptimal new propulsion technology. They discussed the conditions, under which such an event could be avoided and developed indicators for technological and organizational variety. We address the need that exists to escape the current lock-in in the automobile industry first before any possible contemplation of lock-in vis-à-vis another dominant design.

The theoretical concept of lock-in describes a situation, in which an established dominant technological regime blocks all further development and is virtually impossible to replace. Arthur [16] developed a model which shows that, under the condition of increasing returns, a self-enforcement of one technology can emerge, such that all potential users are eventually captured by this technology. In the language of the theory of dynamic systems, one can speak of the existence of an attractor, a stable stationary state which attracts all trajectories in its neighbourhood. What is more interesting from an evolutionary perspective is a situation whereby two attractors – two possible dominant designs – compete. In this case, a situation of hyperselection [24, 25] can arise, in which a technology is highly unlikely to be replaced, even by a much better one. In this paper we use a specific mathematical model of a technological lock-in situation as the reference point to order and evaluate qualitative and quantitative observations in the field of alternative propulsion and fuel technologies. We will first apply the model metaphorically as an analogy to map out the problem; later we will introduce calculations based on this model to validate current replacement experiments in the bus sector in cities and regions.

In reality, we encounter more than just two major competing technologies. The field is more complex. Many propulsion and fuel technology innovations, optimization and substitution processes, incremental and radical innovations exist, which not only push and/or block the paradigm change in very different ways, but also overlap and intersect one another. One of the causes of this confusing and often contradictory situation with regard to innovation arises from the fact that the innovation processes unfold in the crossfire between the pull of convention and innovation pressure. A state of affairs results in which the need for change in the automobile industry increases while the status quo remains on hold. In other words, although all of the stakeholders are aware that traditional propulsion systems cannot survive in the long term, the next possible dominant design has not yet clearly emerged. Different trajectories remain conceivable for the future. As long as different alternatives present themselves as potentially equally successful, competitors hesitate to take the first step. Thus, the core aim of this article is to identify how change can be fostered despite this blockade. We describe the innovation landscape in the area of propulsion and fuel technologies and possible escape trajectories in detail with specific reference to evolutionary theories of technological change.

## 1.2 The pull of convention

The pull of convention emanates from the fossil combustion engine. The history of the second century of the automobile is already being written, however its technical core has remained unchanged. It consists of a combustion engine which imports fossil fuels, in particular oil, from nature and then exports pollutants and greenhouse gases such as carbon dioxide ($CO_2$), methane ($CH_4$) and nitrous oxide ($N_2O$) back to nature. Up to now, the fossil-fuel combustion engine has constituted the convention per se in automobile propulsion technology. In the year 2000, there were around 800 million vehicles with combustion engines in use throughout the world and it is predicted that this figure will have increased to as many as 1.6 million by 2030 [26 (p.2)]. The research and development (R&D) work carried out by the automobile industry remains largely focussed on the optimization of this convention.

The aim of this optimization is to make combustion increasingly efficient and clean. The credo of the engine developers currently bears the rather bold label of "DiesOtto", which pithily expresses the aim of developing an engine that is as efficient as a diesel engine but runs as cleanly as an Otto (petrol-fuelled) engine. A distinction is usually made here between measures for the improvement of conventional functions and downsizing for optimized process management. As ingenious as the wide-ranging DiesOtto optimization is, it is also clear that all of the measures it incorporates are not ultimately directed at substitution but rather at stabilization of the fossil-fuel combustion engine.

The innovations involved in the DiesOtto optimization are Janus-faced and have a fatal inherent logic: they are "Janus-faced" because, on the one hand, every step towards the reduction of dependency on fossil fuels and pollutant substances is urgently necessary while, on the other hand, this optimization stabilizes the dominance of fossil combustion technology. They have a fatal inherent logic because it is unlikely that any technical solutions adopted along this development path will be a "one-stop" solution but will involve mutually supporting auxiliary and partial solutions. To put it in slightly exaggerated terms, what is involved here are upstream and downstream "prosthetic" technologies [27 (pp.87-89)] which may minimize the ecological weaknesses of fossil-fuel combustion technology, but cannot eliminate them.

## 1.3 Innovation pressure

The dominance of the conventional combustion engine is, however, increasingly problematic and it has been subject to growing innovation pressure in recent years. This pressure accrues from three problem fields [28 (pp.8-19)].

- First, the availability and price of fossil combustible fuels. Fossil fuels are becoming increasingly scarce and, therefore, more expensive. Apart from a few exceptions, expert predictions assume that oil and gas prices will continue to increase. It is no longer contested whether prices will rise in the future but by how much. Guaranteeing the long-term availability of oil involves the promotion of so-called non-conventional oil reserves. These include oil sand, polar oil, deep-sea oil (below 500 m), heavy fuel oil and natural gas liquids (NGLs). However, it is questionable and remains to be clarified whether the harnessing of these reserves is economically viable and whether they display a positive general $CO_2$ balance. What is certain is the fact that the times of cheap conventional oil are irretrievably past and the harnessing of non-conventional reserves is highly dubious from an economic and ecological perspective.
- Second, the societal significance of the automobile. The car industry is one of the main pillars of business. It is estimated that in 2005, over 766,000 people were directly employed in the automobile industry in Germany, approximately 1.4 million in the upstream and downstream sectors and a total of 5.3 million people were involved indirectly in the industry. When the car industry sneezes, the economy catches a cold. Thus, automobile innovations are vitally important in both economic and societal terms.
- Third, the pollutant emissions and their consequences. Today, there is no doubt as to the fact that the increase in temperature caused by greenhouse gas emissions will result in considerable changes in the earth's climate. It is also largely uncontested that the reduction of greenhouse gas emissions, in particular $CO_2$ emissions, is urgently needed. This concerns all sources of $CO_2$, including the conventional combustion engine. In addition to carbon dioxide, combustion processes also produce the "traditional" air pollutants. These include, in particular, nitrogen oxide and soot particles. These have a direct impact on human health.

The developments in these three problem fields have resulted in the enactment of legislative regulations and standards which promote and enforce the abandonment of conventional combustion technology. In addition to the European standards Euro 5 and Euro 6, which are due to come into force in September 2009 and September 2014, respectively, the legislative provisions enacted in California, in particular, have fulfilled a pioneering function for the entire automobile industry [27 (pp. 16-26), 29].

The increasing innovation pressure and the legislative regulations and standards, in which this pressure was legally processed and consolidated, resulted in the development of a wide-ranging field of propulsion and fuel technology innovations over the past decade, both in relation to the substitution of the propulsion and the substitution of the fossil fuel. Although these innovations have not yet led to a breakthrough that would prompt a complete departure from the current dominant design, they contain the possible seeds of new technologies. We propose to order these possibilities and develop a typology that enables the assessment of their technical potential for substitution and their survival probability in the context of market competition.

## 2. Innovations: typology and matrix

This very confusing area of propulsion and fuel technology innovation processes can be systematized on the basis of an interdisciplinary innovation typology and an innovation matrix derived from this typology. The different technological options combined in the innovation matrix reflect the current state of knowledge about possible propulsion and fuel innovations. They have been identified using desktop methods (literature review) and interviews in the field. The following sections detail the systematization of these empirical findings.

### 2.1 Innovation typology

The innovation typology is based on one fundamental distinction and three main differences [28 (pp. 25-27)]. The fundamental distinction refers back to Schumpeter and concerns the separation of "invention" and "innovation". An innovation differs from an invention in qualitative terms [30]. The latter denotes mere discovery or development, or to reduce it to a formula: "innovation = commercialization of invention" [31 (p. 328)] or "innovation = invention + exploitation" [32 (p. 3)].

The first main difference concerns the innovation paradigm. Based on studies on the character of the innovation processes in the automobile industry and the term "stagnovation" coined in that context [33], it is possible to differentiate between stagnovative and non-stagnovative innovations. The stagnovative innovations include all innovations which simply optimize the technical core of the automobile, i.e. the conventional technology of the combustion engine. Non-stagnovative innovations are those that do not stabilize the paradigm of traditional thermal propulsion but work on the level of fuel and/or propulsion.

The second main difference concerns the degree of innovation: a distinction is made here between incremental and radical innovations. While incremental innovations involve small continuous developments, radical innovations involve major developments that involve a complete change of direction. Incremental innovations are improvement innovations, radical innovations are significant direction-changing fundamental innovations; or, expressed in more extreme terms, incremental innovations are "Innovatiönchen" (i.e. "minor innovations") [34] and radical innovations are major leaps in an imagined innovation space (sometimes referred to as "quantum leaps") [35, 36].

The third main difference concerns the scope of innovations: we refer here to the work carried out on innovation frameworks [37], system innovations [38], and architectural innovations [65]. We differentiate between modular and systematic innovations. Modular innovations concern propulsion or fuel technology issues solely in relation to the combustion engine of the individual motor vehicle. Systematic innovations extend beyond this into the entire upstream production chains [28 (p. 20)].

The different types of innovation defined on the basis of these three main differences are not unrelated. Many of the studies on innovation theory identify links between the three types, in particular between incremental, radical, modular and systematic innovations. This arises from different conceptual perspectives and is also associated with different terminologies. Thus, these innovation types are often related to each other via various double-field matrices [39 (p. 97)], four-field matrices [37 (p. 8), 40 (p. 97)] and multi-field matrices [41 (p. 9)]. Based on these considerations, four lines were drawn between the innovation types, which enable a further differentiation and systematization of the previously developed innovation typology (figure 1).

**1st and 2nd Degree Innovations**

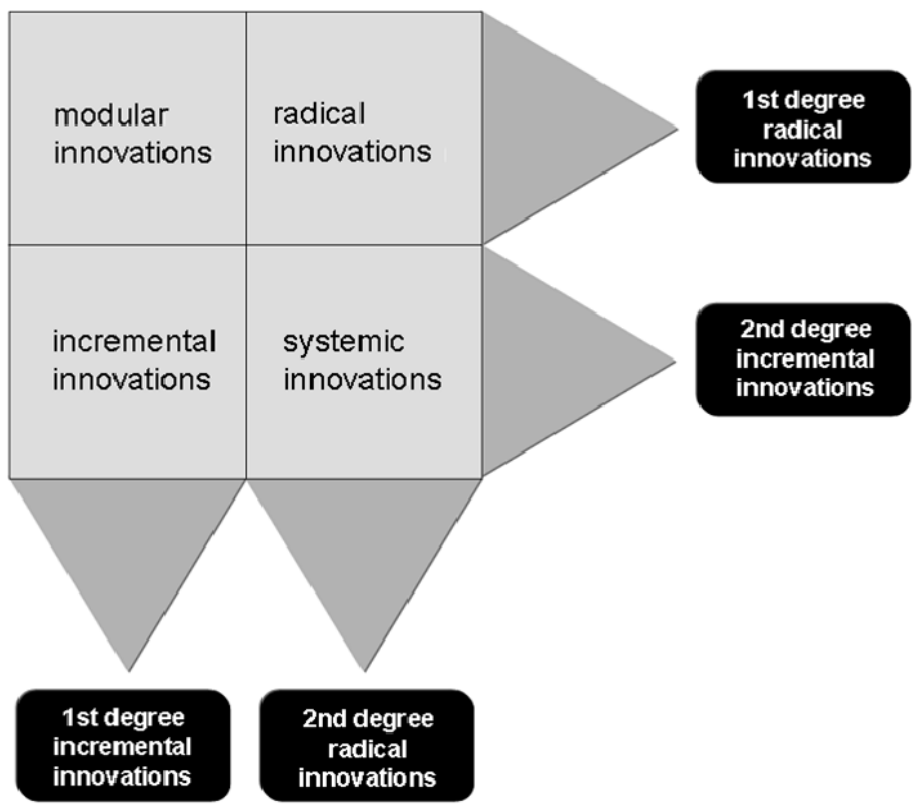

*Figure 1. 1st and 2nd Degree Innovations: The systematics of innovation as a recombination of radical versus incremental and modular versus systemic innovations.*

In the above figure four cells represent the following characteristics of an innovation: incremental, radical, modular and systemic. The notions "radical" and "incremental" describe the technological distance between the innovation and its forerunner (the degree of improvement). The notions "modular" and "systemic" describe the range of influence of a certain innovation on the whole production network (the impact of improvement).[1] These four characteristics can be combined. First degree incremental and radical innovations only have a limited, i.e. modular, scope. They merely concern the technology of an individual vehicle. Second degree incremental and radical innovations, on the other hand, have a bigger and systemic scope. They concern not only the individual vehicle but, beyond this, the entire propulsion and/or fuel technology production chains. However, both modular (1st degree) and systemic (2nd degree) innovations can be either incremental or radical.

**2.2 Innovation Matrix**

Based on the above-developed innovation typology, it is possible to establish innovation matrixes which can enable us to evaluate and systematize the broad and convoluted field of propulsion and fuel-technology innovations. These innovation matrixes were developed within the framework of a research project funded by the German Federal Ministry of Education and Research. This project, 'Alternative fuel technologies in the automobile industry – the sociotechnological coordination of a radical innovation' was carried out at the Social Science Research Center Berlin (WZB) from 2006 to 2008. In the course of the project a profound literature review has been conducted and internal documents of industry-near research institutions and companies in the car industry sector have been analysed. A number of around 40 interviews have been conducted with experts in Germany, China and USA. These more formal interviews have been complemented with a large number of informal expert meetings. On the basis of this combination of standard, codified knowledge on innovation in the area of drive technologies and tacit, embodied knowledge from the experts, different variants for innovations have been identified and ordered in the innovation matrix by two of the authors. [27, 28] The fundamental idea of these innovation matrixes is to coordinate the two main innovation axes in the optimization and substitution of the technology of fossil combustion engines, i.e. the propulsion and fuel-technology innovations axes, as is illustrated in figure 2.

---

[1] Please note that we use "systemic innovation" here with reference to the system of car production. We understand under "systemic" a structural and fundamental change of the whole system under consideration. For a more general concept of systemic innovation see Elzen et al. [42].

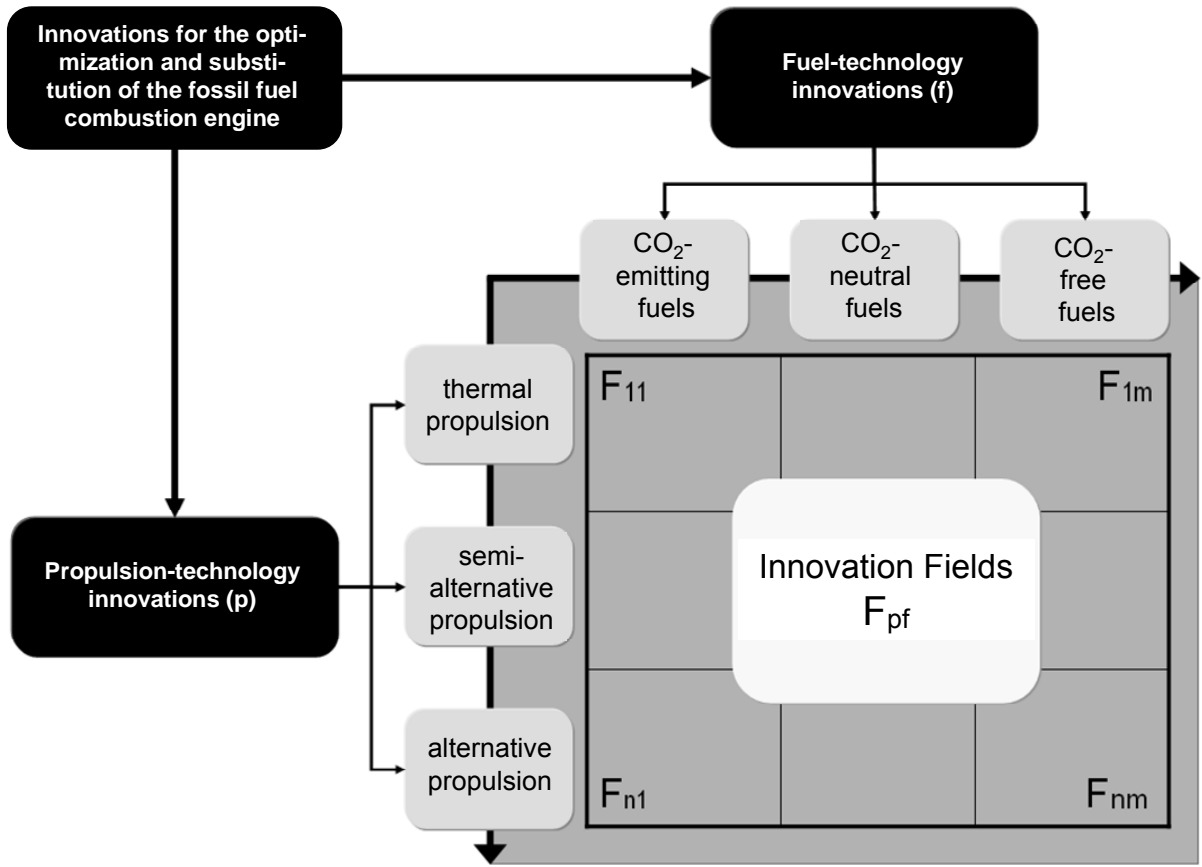

*Figure 2. Fuel-technology and propulsion-technology innovations.*

A two dimensional innovation space with different innovation fields $F_{pf}$ is spanned through the propulsion-technology axis *p* and fuel-technology axis *f*, whereby the index *p* specifies the position of the innovation on the fuel-technology axis and the index f specifies the position of innovation on the fuel-technology axis. The matrix is a discrete typology. It is not possible to specify directly the gap between the different innovation types, whether they may develop out of each other and how many innovation steps lie between them. However, the dimensions can be interpreted in a spatial sense: i.e. different fuels ordered along the imagined *x*-axis in accordance with (decreasing) $CO_2$ emissions and different propulsion technologies ordered along the imagined *y*-axis in accordance with the technological distance away from the traditional combustion engine. This technological distance is meant as a qualitative distance. In terms of the idea of a technological space, like different types of mutations in biological evolution, incremental and radical innovations can be interpreted as smaller or larger jumps. Thus the question can then be addressed as to the "optimal step width" along a technological trajectory [36] and the optimal strategies for the introduction of new technological variants to a market. We will return to this analogy later.

Based on this system, the propulsion-technology and fuel-technology innovations can be compressed into the innovation matrix shown in figure 3 below. A vehicle configuration $F_{pf}$ (field) is defined by the two innovation dimensions *p* and *f*. The *p* index specifies the position of the configuration on the propulsion-technology axis and the index f specifies the position of this configuration on the fuel-technology axis. The $F_{pf}$ fields run from $F_{1,1}$ (row 1, column 1) through $F_{1,15}$ (row 1, column 15) and $F_{8,1}$ (row 8, column 1) to $F_{8,15}$ (row 8, column 15). Seven different configuration levels can be identified with the help of this innovation matrix:

- configurations which are technologically impossible, for example a four-stroke piston engine that runs on electricity ($F_{1,13}$) or a battery electric drive that runs on diesel ($F_{7,1}$). These fields are marked with an "x".
- invention spaces, i.e. configurations which are technologically possible but have not yet reached the innovation stage. These include, for example a GTL-powered Wankel engine ($F_{3,6}$) or Stirling engine that runs on biodiesel ($F_{4,8}$). These fields are marked a dot (•).
- stagnovative innovations, i.e. innovations which do not overcome conventional vehicle configurations in terms of either propulsion or fuel-technology, but merely optimize them. These include, for example, fields $F_{1,1}$, $F_{1,2}$, $F_{2,1}$ and $F_{2,2}$.
- first degree incremental innovations, i.e. innovations which overcome conventional vehicle configurations either on the basis of propulsion or fuel technology but which are only modularly and not systemically innovative. These include, for example, fields $F_{1,3}$, $F_{1,6}$, $F_{2,3}$ and $F_{2,6}$.
- second degree incremental innovations, i.e. innovations which overcome conventional vehicle configurations either on the basis of propulsion or fuel technology and, in addition, are not only modularly innovative but also systemically. These include, for example, $F_{1,7}$, $F_{1,8}$, $F_{2,9}$ and $F_{2,11}$.
- first degree radical innovations, i.e. innovations which overcome conventional vehicle configurations both in terms of propulsion and fuel technology but which are only modularly and not systemically innovative. These include, for example, $F_{7,13}$, if the electricity used is not generated regeneratively, and $F_{8,14}$ and $F_{8,15}$, if the hydrogen used is not produced regeneratively.
- second degree radical innovations, i.e. innovations which not only overcome conventional vehicle configurations both in terms of propulsion and fuel technology but are also modularly and systemically innovative. These include, for example $F_{8,1}$, as well as $F_{8,14}$ and $F_{8,15}$, specifically when the hydrogen involved is produced regeneratively.

Figure 3

| Propulsion (p) \ Fuel (f) | | | CO$_2$-Emitting Fuels | | | | | | CO$_2$-Neutral Fuels | | | | | | C-Free Fuels | | |
|---|---|---|---|---|---|---|---|---|---|---|---|---|---|---|---|---|---|
| | | | Conventional | | Non-Conventional | | | | Semi-Alternative | | | | | | Alternative | | |
| | | | Hydrocarbons | | | | | Synthetic Fuels | Oils | | | Alcohols | | Gases | | Hydrogen | |
| | | | Diesel (1) | Petrol (2) | Autogas (LPG) (3) | Natural gas LNG (4) | Natural gas CNG (5) | GTL (6) | BTL (7) | Bio-diesel (RME) (8) | Vegetable oils (9) | Bio-ethanol (10) | Bio-methanol (11) | Biogas (12) | Electricity (13) | LH$_2$ (14) | CGH$_2$ (15) |
| Thermal propulsion | Conventional | Four-stroke piston engine (1) | | | | | | | | | | | | | X | | |
| | | Two-stroke piston engine (2) | | | • | • | • | • | • | • | • | • | • | • | X | • | • |
| | Non-conventional | Wankel rotary engine (3) | • | | • | • | • | • | • | • | • | • | • | • | X | • | |
| | | Other combustion principles (4) | • | • | • | • | • | • | • | • | • | • | • | • | X | • | • |
| Hybrid propulsion | Semi-alternative | Mild hybrids with battery (one-mode) (5) thermal | | | | | • | • | | | | • | • | • | • | • | • |
| | | Mild hybrids with battery (one-mode) (5) electric | | | | | | | | | | | | | | | |
| | | Full hybrids with battery (two-mode) (6) thermal | | | | | • | • | | | | • | • | • | X | • | • |
| | | Full hybrids with battery (two-mode) (6) electric | | | | | | | | | | | | | | | |
| Electrical propulsion | Alternative | Battery electric drive (7) | X | X | X | X | X | X | X | X | X | X | X | X | | X | X |
| | | Fuel-cell electric drive (8) | • | • | • | • | • | • | • | • | • | • | | • | X | | |

Legend: Stagnovative innovations | 1st degree incremental innovations | 2nd degree incremental innovations | 1st degree radical innovations | 2nd degree radical innovations | X Technologically impossible | • Invention spaces

*Figure 3. Innovation matrix for propulsion-technology and fuel-technology innovations based on own empirical research and presentation design.*

Based on this systematics, not only single or double but indeed multiple innovation levels exist for each of the hybrid configurations ($p = 5,6$), depending on whether the combustion engine or an electrical component of the drive is considered.

The innovation matrix demonstrates very clearly the central role played by the above-described "pull of convention" in the automobile innovation landscape. This "pull of convention" with its "DiesOtto" optimization continues to dominate innovation activity and is the main starting point and objective of R&D work in the automobile industry.

This problem can also be visualized by mapping the innovation matrix onto an innovation landscape (figure 4). With this step we interpret the different categories (propulsion and fuel) as axes of a space of technological characteristics. The third dimension indicates the number of occurring prototypes of cars in the different cells of the innovation matrix based on empirical, qualitative observations. Although, this diagram is an illustration rather than a measured map, it unlocks a line of thought for the measurement and modelling of innovation landscapes. Actually, this visual illustration has been triggered by conceptualizing technological evolution analogously biological evolution or evolutionary startegies. We explain this analogy further in the next section.

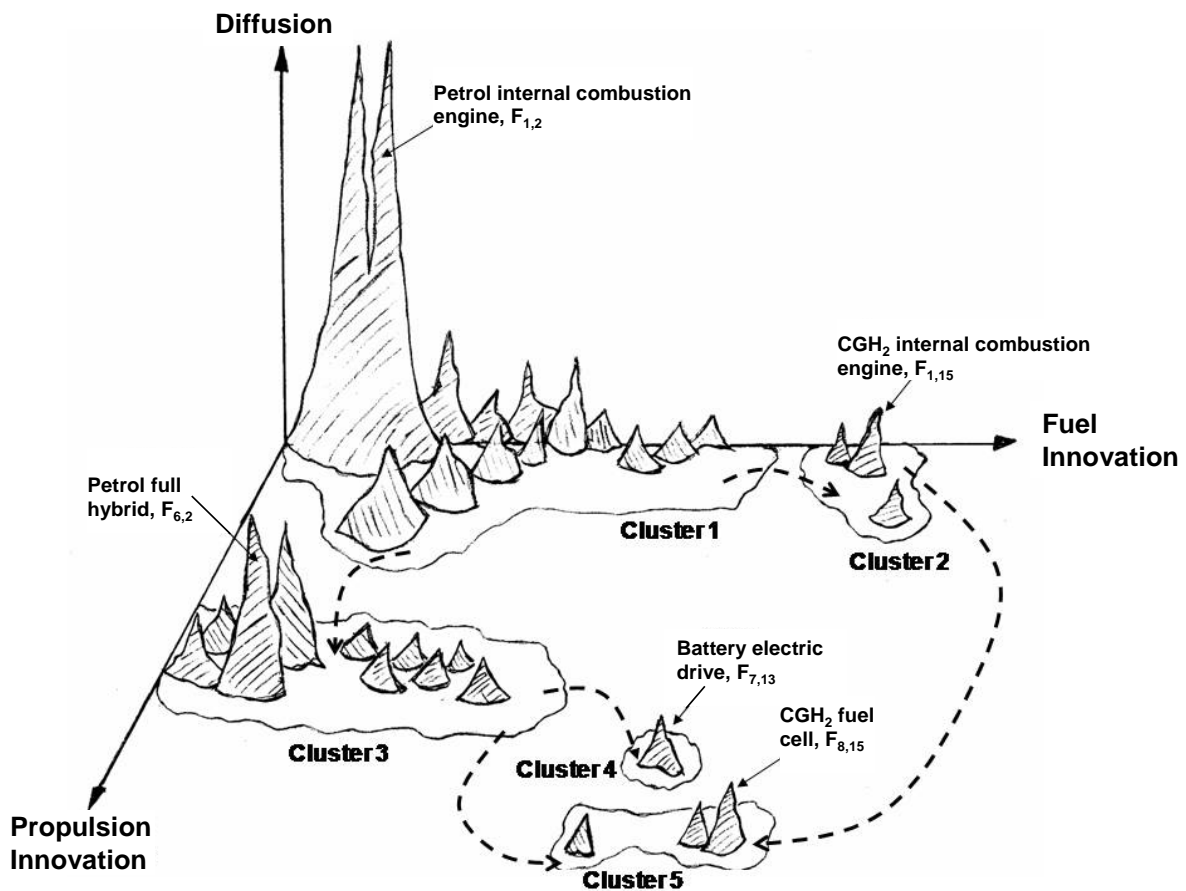

*Figure 4. Illustration of the innovation space around drive technologies.*

Two questions arise from the mapping of this landscape: first, is it possible not only to describe the phenomenon "pull of convention" metaphorically but also to capture it accurately

– 13 –

in conceptual terms? Second, are there ways and strategies for escaping the superior position of the dominant convention and of establishing alternative propulsion and fuel technology innovations? At this point we turn to certain theoretical and mathematical models which have been proposed in innovation theory. We first give a short overview about these approaches. Than, we elaborate on structural possibilities or principle scenarios of survival for a new innovation. The specific model we later rely on in more detail is not new, but what is new is to apply the model not as an explanatory machinery for data sets but as a conceptual skeleton to order qualitative empirical observations.

## 3. Competing technologies in a stochastic model – How to survive in a hyperselective environment

### 3.1 The mathematical model

Since Sahal's [43] spatial representation of technological trajectories as rivers in technological landscapes, the analogy between technological evolution and other dynamic processes governed by a global function to be optimized has been developed in various directions. These include attempts to link technological and biological evolution [44, 45, 46] or to link technological evolution and optimization [47]. However, only a few attempts have been made to visualize this abstract landscape as a landscape with phenomenological properties. The visualization in figure 3 is comparable to measurements and visualization of technological spaces which are based on the analysis of technological and service characteristics by Saviotti and co-authors [48, 49]. Very recently, Alkemade and co-authors [50] proposed a discrete fitness space for alternative vehicles and discuss possible optimal innovation pathways. Stable configurations, such as the use of certain technologies over others, can be visualized as preferable locations in such an abstract space and technological change can be visualized as movements or trajectories towards certain technological configurations [51].

As we know from the study of the processes behind technological change, feedback loops that lead to a self-enforcement of growth are important. This phenomenon was described in Arthur's theory of "increasing returns" [16]. In some cases, this can lead to a situation whereby one technology dominates the entire market (lock-in). In the theory of non-linear dynamics, this phenomenon is known as approaching an attractor. An attractor is a stable stationary state, towards which all possible trajectories of the system that happen to start in its "basin of attraction" (area of influence) develop. However, lock-in or dominant design often refers not only to the emergence of such a preferable situation but also to the apparent immunity of the system to follow-up improvements. It would be assumed normally that a new and better technology will shake up or disturb the system, change the attractor landscape and give rise to new competition resulting in another attractor. A situation whereby the system is bound to one attractor and immune to disturbance is described as hyperselection [24]. Hyperselection can be found in deterministic models; it has also been termed "once-and-forever" selection. For this type of model "a disturbance" corresponds to a new initial condition but further processes are disturbance free. In other works one author has shown that this kind of lock-in situation in technological evolution can be mapped onto the problem of two co-existing (and competing) stable states in a system driven by a special non-linear dynamics [25, 52]. In particular, if permanent disturbance is inherent in the system, stochastic models are needed, but these lead to different results than those obtained from classic hyperselection.

We illustrate the different role of disturbances or fluctuations by examining two possible final states of the system, each surrounded by a "basin of attraction" and separated by a barrier (separatrix) (see figure 5). However, due to the fact that in most cases a new technology will start with only few exemplars, disruption of the previous attractor will be too weak to lead to sustainable change. The system may then stagnate. This seems to be precisely the situation involved in our example of propulsion technologies. Analogous to the convention would be a mother who devours all of her children. In figure 5 we visualize different dynamic strategies which enable the barriers (separatrix) between two competing designs to be crossed. A disturbance of the dominant design (the emergence of a new variant) can be visualized as a moderate deflexion of the system away from its attractor. In other words a new initial condition for the on-going dynamic processes is created. As long as this disturbance is still in the basin or sphere of attraction of the old technology, the system driven by inner forces will usually return to the initial attractor. Successful intervention strategies will be based on fluctuations which drive the system across the barrier (case a) or which create an initial condition far away enough from the old basin (case b). Case c illustrates how re-sizing the system (creating a niche) automatically increases fluctuations (according to laws of probability) and, in turn, increases the likelihood that the barrier will be transgressed. Such strategies can be mathematically calculated using an analogy to non-linear dynamic systems. Very early "escape" strategies which take advantage of the role of fluctuations or disturbances in non-linear systems have been proposed for the case of technological evolution [53, 54].

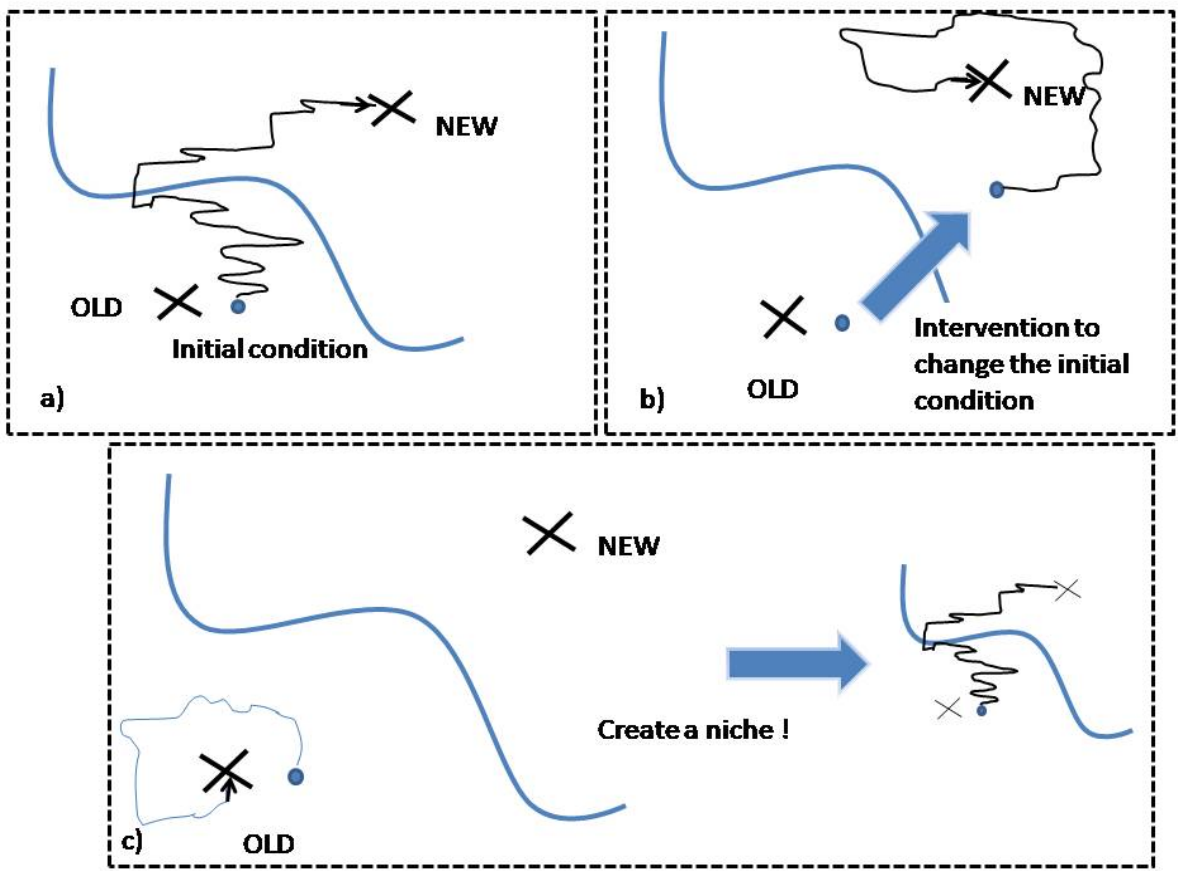

*Figure 5. This is a schematic illustration of the course of a technological trajectory between two different dominant technological designs. Each design is assumed to be represented by a (point) attractor.*

In sum, then, large fluctuations can push trajectories to cross the boundary between the two distinct attractor states (figure 5, a); this has been demonstrated mathematically. Such large fluctuations could also be caused by external shocks to the system or by internal persistent instability, such as uncertain user behaviour. A second option involves restarting the dynamics with a large number of adopters of the new technology (figure 5, b). In so doing, the system is "artificially" taken out of one attractor basin and moved into the other. A third way is to create a niche for the new technology [25] (figure 5, c). Even small fluctuations in this niche can push the system towards the new state.

However, in all cases, a stochastic description is required to describe these effects. The corresponding model and its economic interpretation have been presented elsewhere [25]. In this paper we start from one main outcome: i.e. the survival probability of a new technology. Please note that we use a model that represents just one important feature of technological evolution, i.e. the transition from one technology to another one. Disregarding the multi-state, multi-level and multi-actor character of the process, we chose an "archetypical situation", i.e. the transition between just two technologies with very clearly defined growth potential. The growth potential (or, in mathematical terms, the growth rate) is an expression of the "quality" or fitness of the technology in this simple model. We assume that in a situation of infinite resources, infinite possible users and no competition, both technologies would grow. The difference in these potential growth rates, expressed by the ratio of the new growth rate to the old growth rate, represents the improvement brought about by the new technology. A limited market size introduces competition, and the decision processes of adopters are modelled as transitions from the new to the old technology and vice versa. In the case of a finite market, each growth step for one technology is accompanied by a decline step for the other technology. Thus, from the processes of entry in, exit from and transitions between technologies only transitions are left. The transition rates are equal to the growth rate in an unlimited market. The improvement (or ratio between the transition rates) is also referred to as the selection advantage of the new technology.

How the technologies would grow makes a significant difference in the dynamics. If we assume that the rate of growth is constant, the result is linear growth. A feedback loop of the first order (or linear growth rate) leads to exponential growth over time. In this case the growth rate depends on the number of users of the technology. A feedback loop of the second order (or quadratic growth rate) leads to hyperbolic growth and hyperselection [55]. The model we use [25] covers all of these cases. In this paper we focus on hyperselection (quadratic growth). For our archetypical situation a survival probability function for the new technology (alternative) can be calculated analytically. The survival probability of the old technology (conventional) added to that of the new technology equals one. In other words the survival probability of the new technology is equal to the extinction probability of the old one. Therefore, we need only consider the new technology.

We have illustrated how hyperselection can be broken up in the presence of fluctuations. In the deterministic case, independent of any improvement, the survival probability of a new technology would simply be zero. In the stochastic model we obtain a formula that provides probability values for survival based on only three parameters: the size of the market $N$, the initial condition (starting number of users) for the new technology $N_A$ and the selection advantage of the new technology $Q$. If $Q$ is greater than 1, the new technology is better than the old one. In the case of quadratic growth the formula is:

$$\sigma(N, N_a, Q) = \frac{1 + \sum_{i=1}^{N_a-1} \left(\frac{1}{Q}\right)^i \times \binom{N-1}{i}}{\left(1 + \frac{1}{Q}\right)^{N-1}}$$

(1)

In this equation $N_A$ is the initial number of users of an alternative technology, $N$ the number of all users in the system (the size of the market), $Q$ the ratio between the growth rate $W_A$ of the new alternative technology $A$ and the growth rate $W_C$ of the old conventional technology $C$ and the survival probability $\sigma$ of the new technology. This model enables us to devise a series of ideal-type simulations (numerical explorations of equation 1) from which strategies for overcoming hyperselection can be derived.

## 3.2 The simulations

To explore the theoretical framework the model presents as basis for ordering empirical observations we have carried more simulations than presented with the first publication of the model [25]. We visualize how dependent the survival probability $\sigma$ of the alternative technologies is on the parameters $N$, $N_A$ and $Q$. In order to enable the graphical presentation of $\sigma$ in a two-axis coordinate system, two of the three variables must be set as constant so that σ can be plotted against the remaining third free variable. The definition of certain sets of parameters enables the variation of the latter and sets of curves are generated. For example, if $N$ is defined as a constant and $N_A$ varied across three specific values, a set of curves arises over $N_A$ with three curves in a function graph, in which σ is plotted against $Q$. Thus the following possible simulations of survival probability are conceivable in principle:

|  | σ against Q | | | σ against N | | | σ against $N_A$ | | |
|---|---|---|---|---|---|---|---|---|---|
|  | N const., $N_A$ var. | N var., $N_A$ const. | N var., $N_A$ const. | Q const., $N_A$ var. | Q var., $N_A$ const. | Q var., $N_A$ var. | Q const., N var. | Q var., N const. | Q var., N var. |
| Linear cases | X | | | | | | | | |
| Quadratic cases | X | X | X | | | | | | |

Const. = constant; var. = varied across a certain set of constants.

*Figure 6. Table of functional dependencies between three parameters. The exes (x) mark those cases which are dealt with in detail below.*

We will now present four of the 18 possible simulations by way of example (indicated in figure 6 by an x). All four simulations are carried out for the case "σ against $Q$" which means that the results display a good degree of comparability. Only one simulation is carried out for the linear case, i.e. simulation 1, whose parameters are indicated by an asterisk (*). The equation for this case can be found in [25]. When these simulations are compared with the quadratic case, what emerges very quickly is that the linear case represents a fundamentally different situation from hyperselection. Although we have defined hyperselection as based on quadratic growth, hyperselective tendencies understood as temporal dominance of the old technology can be also found in models with other growth modi, e.g. linear growth. It is worth mentioning that the survival probabilities need to be understood as survival probabilities in the long run. In the short run a conventional technology can remain persistence just due to the time required (although avoidable) for a transition from the old to the (apparent temporary virtual hyperselection). On the other hand, even in a situation of hyperselection, a new technology can persist in the system for a while (apparent temporary virtual survival). For the interpretation of the model, therefore, it is important to remember that our statements refer to the conditions of sustainable technological change in the long run. The following four individual simulations are carried out:

- simulation 1, where σ* = *f(Q\*)*, $N_A$* is varied and $N$* is constant (linear case);
- simulation 2, where σ = *f(Q)*, $N_A$ is varied and $N$ is constant (quadratic case);
- simulation 3, where σ = *f(Q)*, $N$ is varied and $N_A$ is constant (quadratic case); and
- simulation 4, where σ = *f(Q)*, $N_A$ is varied and $N$ is varied (quadratic case).

The first two simulations demonstrate very clearly how strongly dependent survival probability σ is on the parameter $N_A$ in both the linear case (simulation 1) and the quadratic case (simulation 2). Figure 7 shows first the result for the linear simulation. In this graph, $\sigma^*$ is represented as a function of $Q^*$, $N^*$ is constant ($N^* = 20$) and $N_A^*$ varied across the parameter set ($N_A^* = 10$, $N_A^* = 8$, $N_A^* = 5$, $N_A^* = 3$ and $N_A^* = 1$).

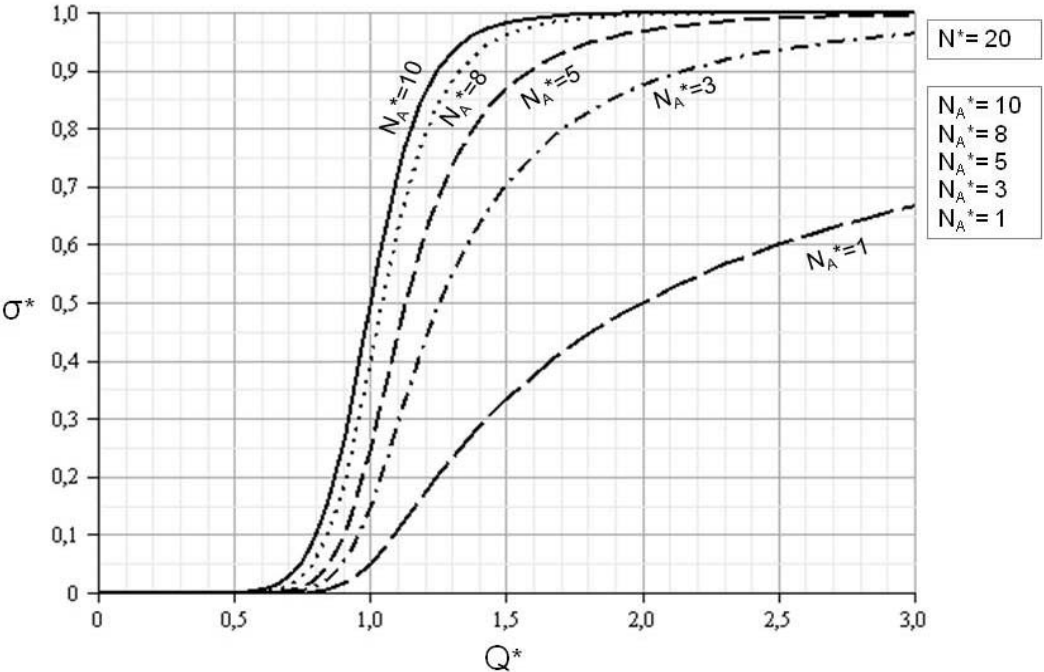

Figure 7. Simulation 1: $\sigma^* = f(Q^*)$, $N_A^*$ is variable and $N^*$ is constant (linear case).

The simulation shows, that survival probability $\sigma^*$ declines with a diminishing $N_A^*$. Even if $Q^*$ is greater than 1, i.e. the growth rate $W_A^*$ exceeds growth rate $W_C^*$, the survival of the alternative technology is not certain and only comparatively small. This differs importantly from the deterministic model (without fluctuations) in which – in the linear case – the better technology always wins. The set of curves shows, that in the linear case with a relatively small $N_A^*$ for the alternative technology A, the latter has very good chances of survival against the dominant technology C even if $Q^*$ is smaller than 1. As soon as technology A grows just slightly faster than technology C, or $W_A^*$ is just slightly higher than $W_C^*$, $\sigma^*$ increases rapidly. This is demonstrated clearly by the curve for $N_A^* = 3$: if alternative technology A grows at a rate approximately 30% faster than the conventional technology C ($Q = 1.3$), the survival probability for A is already 50%. If $N_A^* = 5$, it actually increases to over 70%. If $Q$ is less than 1, the survival probability of A is very low – however, unlike the deterministic model, it actually still has a chance of survival.

Exactly the same parameters as those used in the linear case, are now selected for simulation 2. The following graph results (figure 8).

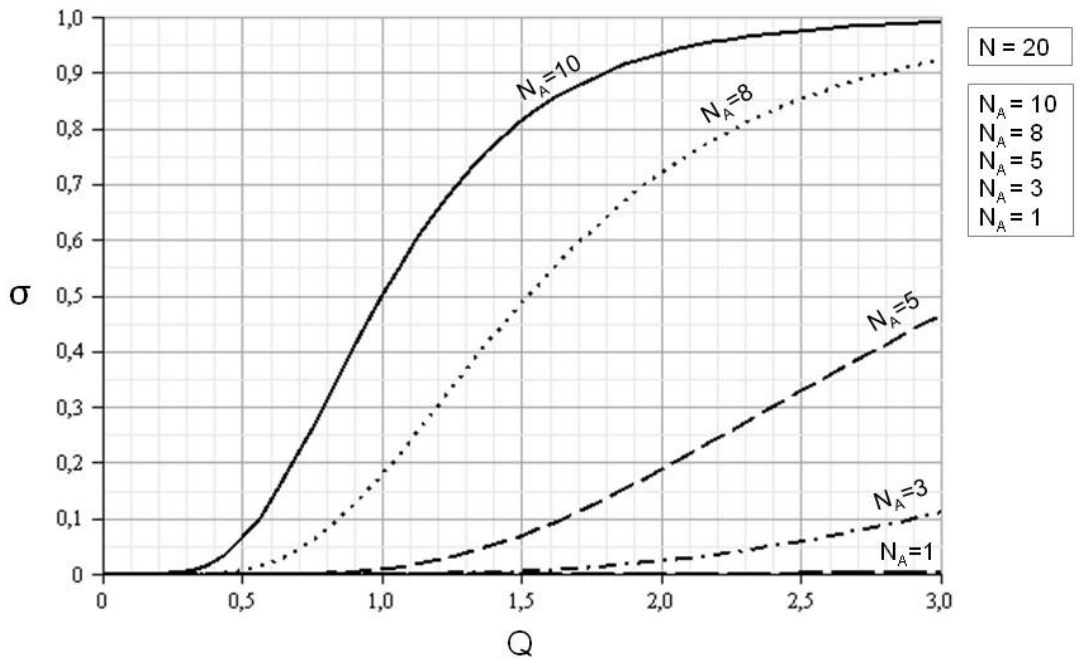

*Figure 8. Niche simulation 2: σ = f(Q), $N_A$ is variable and N is constant (quadratic case).*

The hyperselective behaviour of the conventional technology emerges particularly clearly in the quadratic case. Unlike the case in simulation 1, the alternative technology *A* must attain significantly higher growth rates here to be able to survive against the conventional technology. This is reflected in the fact that the curves in the quadratic case are clearly flatter than in the linear case. It is clear from the graph that, in contrast to the linear case, small batch sizes ($N_A < 3$) have practically no chance of survival. Even if $Q = 2$, the survival probability at $N_A = 5$ falls to 20% and as far as 2% at $N_A = 3$. For the case $N_A = 1$, σ is almost 0%, even if growth rate $W_A$ is three times higher than $W_C$. A comparison of simulations 1 and 2 shows that the linear model does not reflect hyperselective effects sufficiently and is still too "deterministic" in the sense that a new, better technology almost always wins.

The third niche simulation shown in figure 9 clearly demonstrates how in the quadratic case survival probability σ depends on the parameter *N*. σ is also presented as a function of *Q* here but in this case $N_A$ is set as constant ($N_A = 1$) and *N* is varied across the parameter set $N = 2$, $N = 10$, $N = 20$, $N = 50$ and $N = 100$.

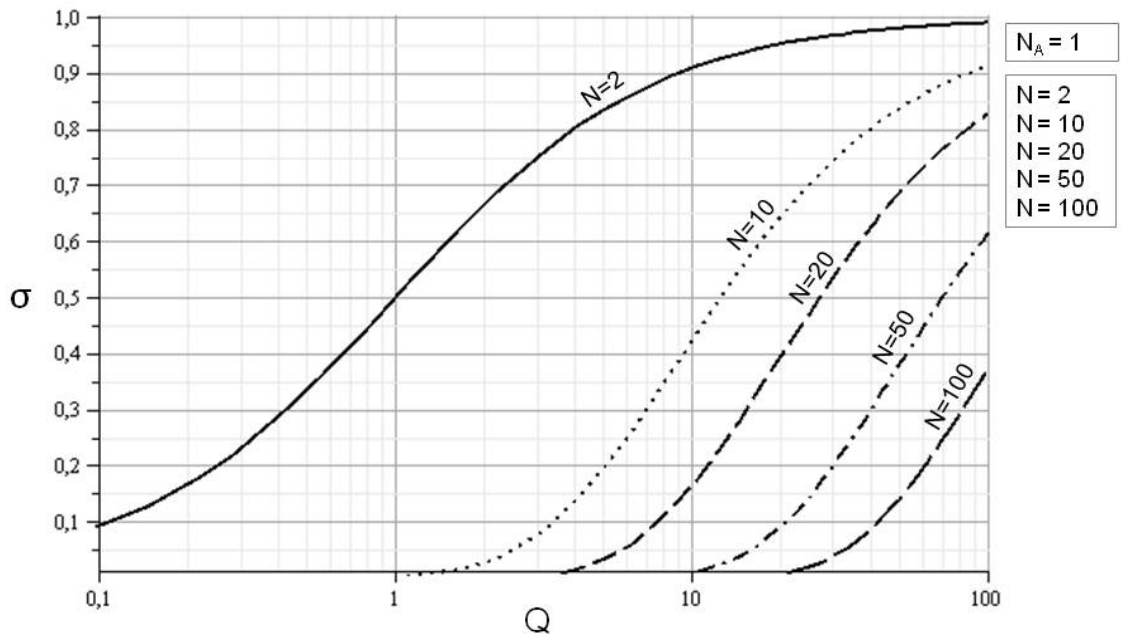

*Figure 9. Niche simulation 3: σ = f(Q), N = variable and $N_A$ = constant (quadratic case).*

The bigger *N* is, the lower the survival probability of alternative technology *A* with constant $N_A$. Even if *Q* is considerably greater than 1, if the growth rate $W_A$ exceeds growth rate $W_C$ by a multiple, the survival probability declines with an increasing *N*. For example, if $W_A$ exceeds $W_C$ by a factor of ten, the survival probability at *N* = 2 is around 90%; as opposed to this if *N* = 10 it declines to just over 40% and to almost 0% at values of *N* ≥ 50.

The two simulations developed from the quadratic case, i.e. simulations 2 and 3, can also be combined with each other so that not just one but two parameters are varied at a certain functional dependency. This is presented exemplarily in figure 10 for the functional dependency *σ = f(Q)*. Both *N* and $N_A$ vary here and in such a way that they are in an equal ratio to each other. The following value pairs were selected: *N* = 100/$N_A$ = 50, *N* = 50/$N_A$ = 25, *N* = 20/$N_A$ = 10 and *N* = 10/$N_A$ = 5. The resulting simulation takes the following form (figure 10).

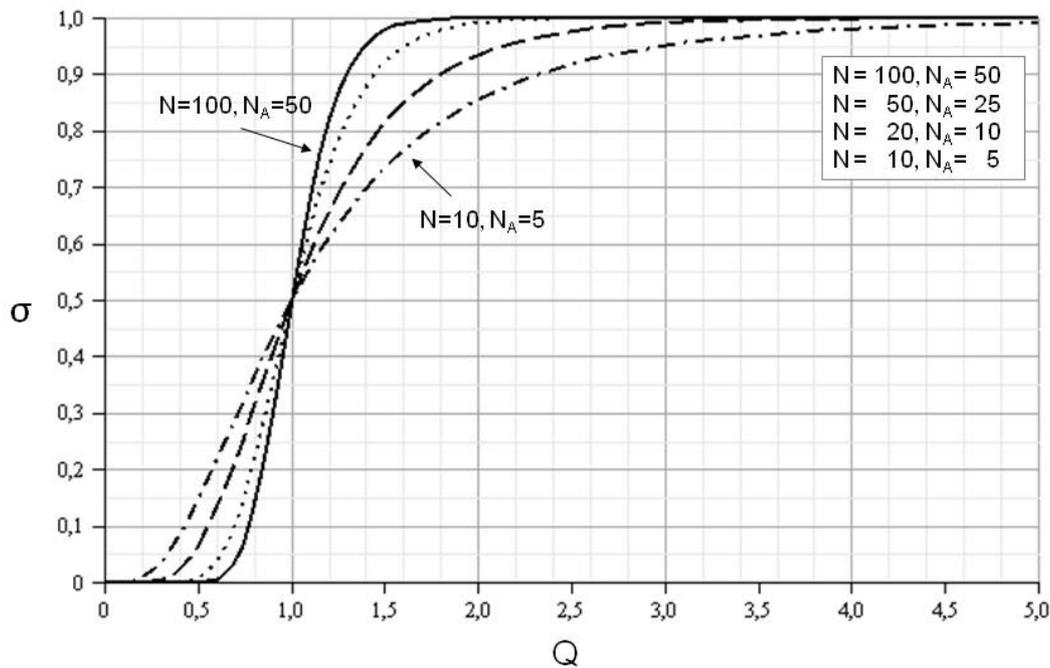

*Figure 10. Simulation 4: σ = f(Q), $N_A$ is constant and N is variable(quadratic case).*

This simulation clearly indicates that the survival probability of alternative technology *A* not only depends on the relation of *N* to $N_A$ but also on the latter's absolute values. If $Q \leq 1$, i.e. $W_A$ is smaller than $W_C$, the survival probability *σ* is higher at smaller absolute values than at bigger ones. For example, at a growth quotient of 0.5, the survival probability for the case *N* = 10 and $N_A$ = 5 is approximately 20% but declines to almost 0% for the case *N* = 100 and $N_A$ = 50. However, if Q ≥ 1, i.e. $W_A$ is greater than $W_C$, the situation reverses. In this case the survival probability is higher at greater absolute values than smaller ones. For example, at a growth quotient of 1.5, the survival probability *σ* is almost 100% for the case where *N* = 100 and $N_A$ =5 0 and is only 70% for the case *N* = 10 and $N_A$ = 5.

    Based on the results of these four simulations, it is possible to formulate the following rule of thumb for the development of innovation strategies: In order for alternative technology to survive in the hyperselective landscape, *N* must be as small as possible and both $N_A$ and *Q* must be as big as possible.

## 4. Application: Simulations and strategies

Despite its non-linear and complex nature, the above-presented model has only three parameters, namely, the size of the system, the initial condition and the degree of improvement. The size of the system is equal to the size of the market, in which the competition operates. This can be measured in terms of potential users or approximated by the number of produced or sold cars in total. The initial condition is the number of users of the new type of automobiles, or the number of new type of cars produced. Measuring improvement is more difficult however. In the model, the improvement is given by the ratio between the two growth rates the technologies would have in an unrestricted market. We can also call this the growth potential. For the linear case, it is the number of new adopters of the new technology per time unit divided by the number of already existing. In the so-called quadratic case the number of new adopters is divided by the square of the number of already existing adopters. While we may not be able to "measure" each of the parameters we will show how such a model based on three parameters can be used to design strategies for the introduction and survival of new technologies.

### 4.1 Strategies

Buses offer good options for fulfilling the criterion for survival: small $N$, big $N_A$ and big $Q$ proposed by the model as optimal as possible. Buses represent a particular niche for two reasons: first, the bus niche is a representative niche and, second, it is also a multiple niche. Figure 11 illustrates clearly how buses present a representative niche.

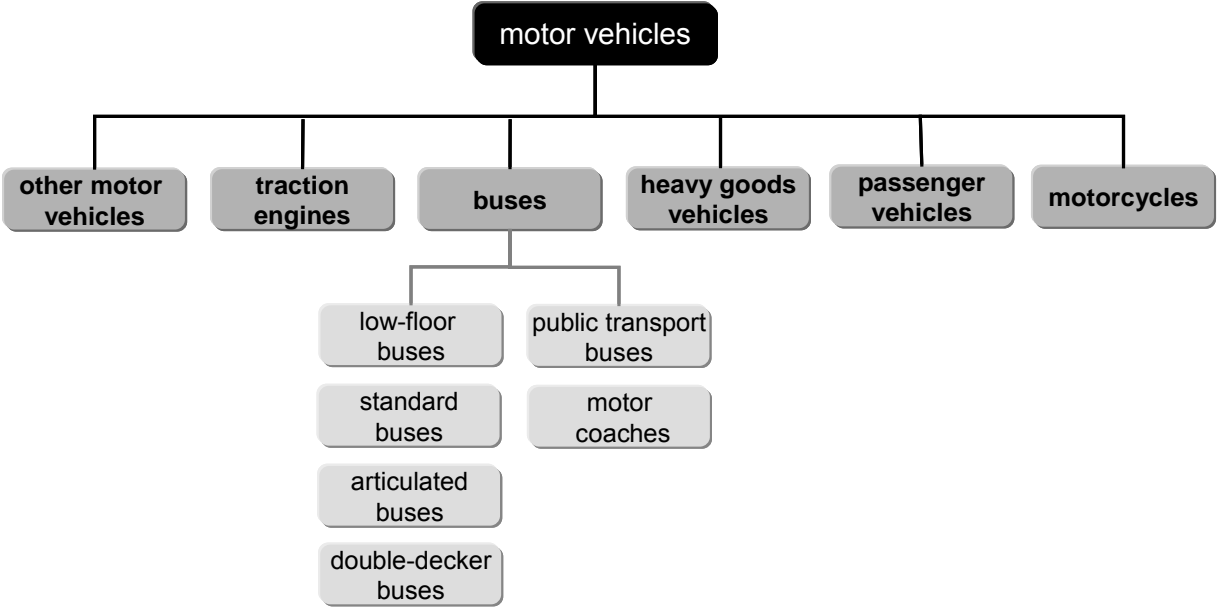

*Figure 11. Classification of vehicles.*

Buses constitute an important sub-group among motor vehicles [56 (p. 41)]. This is illustrated very clearly by the example of Germany.
In Germany, the bus is the second most important means of passenger transportation after the car. It is used for over five billion journeys annually. For short-distance public transport where it accounts for more than half of journeys, it is the number one means of transport. Buses represent the cornerstone of short-distance public transport, in particular in the outskirts of large and medium-sized cities and in rural areas in particular. In the absence of the bus, people would lose a considerable amount of mobility. Germany's stock of 82,600 short-distance transport buses contrasts with 9,083 urban trains and trams. Buses provide capacity for 6.58 million people as opposed to the 1.17 million places provided by the remaining short-distance public passenger transport. While urban rail and trams cover 620 million vehicle kilometres per year, buses cover 3.18 million, i.e. five times more. For a further 120 million passengers, motor coaches also offer a comfortable "large-capacity limousine" for long-distance travel.

Around 6,200 companies are currently involved in short-distance public passenger traffic in Germany, the majority of which are private operations. In total, all of the transport service providers employ 178,000 people, of whom 69% are directly involved in the provision of transport services. Thus, a total of around 750,000 jobs in Germany are directly and indirectly dependent on the bus sector.

The figures clearly demonstrate the enormous economic significance of bus transport in the area of passenger transport and would not initially evoke an impression of a niche phenomenon. The niche nature of bus use only becomes clear through comparison with the economic importance of the car. As compared with motorized individual transport by automobile, short-distance public transport only plays a subordinate role in the overall transport system. While motorized individual transport is used for over 147 million journeys daily, only 18,435 journeys are completed using short-distance public transport. Moreover, at 55.7 million units in 2001, global production of passenger cars was 400 times higher than that of buses. Thus buses are a niche but they are not peripheral, i.e. they constitute a representative niche. Furthermore, this representative niche is also a multiple niche.

Buses represent first a double niche. In addition they constitute an application niche in the area of motor vehicles. For example, a total of approximately 55.5 million motor vehicles were licensed in Germany in 2006. At almost 47 million vehicles, the private passenger car represents the largest segment in the motor vehicle segment by some distance. Only approximately 83,500 vehicles are licensed in the bus segment. Thus, buses merely represent 0.2% of the total motor vehicle stock in Germany.

Buses also constitute a spatial niche. The bus market can actually be understood as a collection of many small niches, within which local transport companies operate their bus fleets. They represent independent organizational units and act in a spatially delimited territory which can be described as a spatial niche. Thus a transport company's bus operation represents a niche within the niche. This double niche forms a particular protection area for innovations which can be visualized as shown in figure 12.

Buses can represent not only a double niche, but also a multiple niche. Thus, for example, additional spatial or functional segmentation, e.g. shuttle transport or sightseeing tours, takes place in the context of an urban application/spatial niche. Such niches then become triple niches, i.e. niches in a (spatial) niche, which in turn is located in an (application) niche.

This formation of niches within niches is a strategy suggested by both the niche simulations and the bus transport innovators, who often develop such strategies intuitively. Multiple niches are multiple protection areas for the survival of innovations. They create the structural condition for survival, i.e. small $N$, big $N_A$ and big $Q$. We have described elsewhere [56] fleet experiments with buses in Europe and overseas which correspond to different fields of the innovation matrix. The following two case studies from the USA are particularly suited to visualizing the role of nested niches as a paradigm.

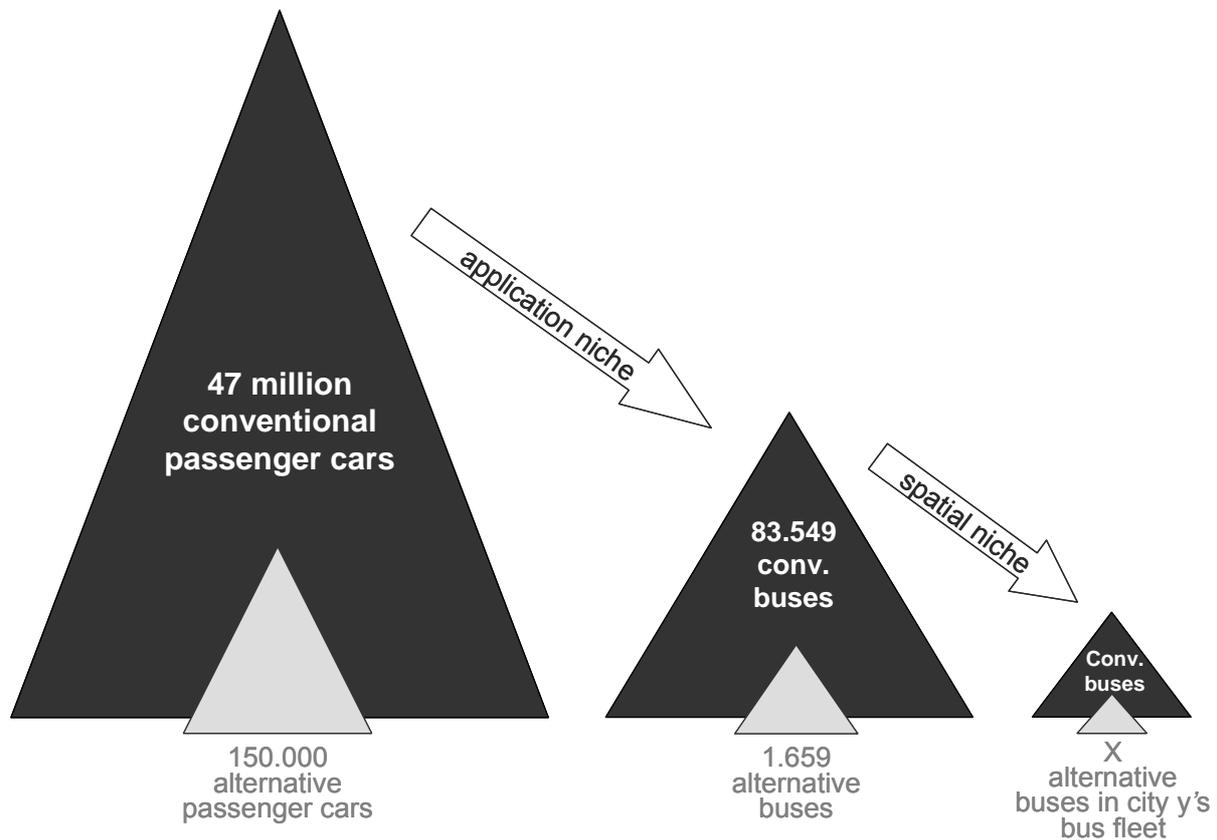

*Figure 12. Double niches as protection areas for innovations.*

## 5. Case studies: SunLine and CARTA

### 5.1 SunLine

The first example involves the company SunLine Transit, whose headquarters are in Thousand Palms in California. SunLine commenced operation in 1977 with 22 vehicles and a network of routes extending across over 2,849 km² in Coachella Valley in California. In 1992, Sunline decided to exchange its entire diesel fleet for CNG (compressed natural gas) buses. The aim was to become the leading transport company in relation to the reduction of emissions. Thanks to its complete conversion to a new technology, SunLine skipped the pre-market stage and directly created an early market for CNG buses. As a result of this complete conversion, $W_C$, the growth rate of the conventional technology, is zero and $N$ equals $N_A$. The hyperselective convention is thus completely eliminated.

This was only possible due to the comparatively small size of the bus fleet. In the case of a large fleet, this conversion would have to take place in stages. However, due to the application niche of the bus market and the spatial nice of the transport company SunLine, in this case the overall size $N$ was comparably small and the new technology could displace the convention in one fell swoop. Thanks to the special characteristics of a double niche, it was possible to rupture the superior position of a hyperselective convention.

The innovativeness of this conversion cannot be estimated too highly from today's perspective. At the time there were only 25 full-size CNG buses in operation in the USA and people had relatively little experience with the new technology. Thus, for example, a new training centre for dealing with alternative fuels was established in 1993 to prepare all of the SunLine employees for the operation of the new buses. The new CNG buses were delivered in 1994 and SunLine was the first transport company in the USA to have a complete fleet of CNG buses. The inevitable teething problems could be quickly overcome through close cooperation with the engine manufacturer and other transport companies, such as the Los Angeles County Metropolitan Transit Authority.

Sunline currently has over 46 CNG full-size CNG buses. The conversion to CNG fuel has been deemed a success. Thus, SunLine plans to replace its entire fleet again with new CNG buses in 2009. This clearly demonstrates once again that what is involved here is an early market. As although the complete fleet now consists of CNG buses, these must still prevail over the convention to avoid a change back to the diesel buses. Thus several scenarios are conceivable for the future. The CNG buses could, of course, prevail in the long term and SunLine will continue to operate its buses on natural gas. However, a return to diesel would also be possible. The gradual replacement of the CNG buses with an even more innovative configuration is also conceivable.

Based on this, other propulsion and fuel technologies are already being tested at SunLine. For example, the company is involved in hydrogen projects. A hydrogen-powered fuel cell bus was tested in 2001. Another fuel cell bus was in operation at SunLIne from 2002 to 2003. Buses are also being tested which are run on a mixture of hydrogen and natural gas. The aim is to further reduce the nitrogen oxide emissions of the buses and to test a reliable "bridge" on the way to pure hydrogen propulsion. Thus, a hybridized hydrogen bus with combustion engine was introduced by the company in 2004 and a hybridized fuel cell bus has also been in operation since 2005. SunLine has also expressed the desire to test other fuel cell and hydrogen buses if the necessary funding is provided.

This transport company can be viewed as a pioneer in the area of CNG buses in the USA. Under the favourable starting conditions, it succeeded in creating an early market for CNG buses. This example attracted attention and the number of buses in the USA operated using LNG and CNG increased from 25 in 1993 to a total of 3,500 in 2000. The situation would have developed in a different way if the new technology had not had the possibility of surviving against the dominance of a hyperselective convention in the protection of a relatively small spatial niche. Thus the foundation for the success of CNG buses in conquering ever expanding areas in the automobile innovation space was laid in Thousand Palms.

## 5.2 CARTA

The second example involves the transport company Chattanooga Area Regional Transport Authority (CARTA) which succeeded in establishing a triple niche for battery-run buses. Chattanooga is a city in the American Federal State of Tennessee the USA with 153,431 inhabitants. Due to high automobile density and extensive industry, the level of air pollution in the city was extremely high. In 1969, Chattanooga gained the dubious honour of being the US city with the highest level of diesel soot particles. The city reacted by passing a law for the reduction of air pollution by industry. Companies complied with the new standards, which resulted in a clear improvement in the city's air quality. However, the problem of automobile density remained and up to 1989 Chattanooga still had the reputation of being one of the most polluted cities in the USA.

In order to reduce the traffic in the inner city, CARTA proposed to set up a shuttle bus service between the south and north boundaries of the city centre. Parking would be provided at both terminuses so that people could park their cars there and take the bus. To further reduce pollutant emissions, it was proposed to use emissions-free battery buses for the shuttle service. This represented a third niche in the application and spatial niches. The total size $N$ became increasingly small as a result. The additional niche in this case was implemented with the help of subsidies. Subsidies of USD 15.7 million granted by the Federal Transit Administration, USD 2 million from the Tennessee Valley Authority and USD 2 million from the Tennessee Department of Transportation made it possible to implement the plans; the parking areas and first battery buses were financed in full by the subsidies. At the same time, the private non-commercial Electric Transit Vehicle Institute (ETVI) was established which was intended to support the use of the new technology through R&D.

The first two battery buses were delivered to CARTA by a manufacturer in California in 1992. The bus manufacturer Advanced Vehicle Systems (AVS) was established in Chattanooga the same year. Shortly after this, CARTA placed an order for 12 battery buses with AVS, which were developed with the support of the Californian bus manufacturer. Despite teething problems, the battery buses are still used in Chattanooga today. It is even planned to order more battery buses in 2008. A total of 25 battery buses are now in operation. The project is viewed as a major success. The buses contributed to an increase in the general use of public bus transport. The passengers appreciate the low level noise pollution and stable handling offered by the battery buses as compared with standard diesel buses.

Thus, an early market for battery buses grew from the shuttle-bus niche in Chattanooga. They still have to prevail against the diesel buses under market mechanisms, but within the niche conditions of the shuttle service. The hyperselectivity of the convention was, therefore, eliminated and the new technology was able to become established in the protection area of a niche. This protection area in form of the shuttle service displays some special characteristics which are of enormous significance for the success of the project. These characteristics include the following specific factors.

- The battery buses were subject to relatively favourable conditions in their operation as shuttle lines cover comparatively small distances as compared with normal urban routes. The shuttle service in Chattanooga merely serves the city centre, which covers a distance of 3.2 km. Moreover, the city centre area is very flat thus the busses do not have to deal with major inclines.
- A new depot was built at the southern terminus of the shuttle line in 1994 at a cost of USD 4.2 million. This included 550 new parking spaces, a battery changing station and battery charging equipment. Operation starts at 6 am and the buses return to the depot at midday and their empty batteries are exchanged for newly-charged ones. Thus a battery bus covers an average daily distance of 161 km.

The increased use of short-distance public passenger transport reflects the success of the project. This is due inter alia to the fact that the shuttle service is provided free of charge. The operation of the buses is financed exclusively by income obtained from the car parks at the bus terminuses. Short distances, the absence of inclines in the inner city, the depot regime and financing through parking fees show how a protective space for new technologies can emerge through the creation of a niche. The battery buses only compete with the diesel buses under very specific conditions. Thus, relatively good operating conditions are created for the battery buses through a niche within a niche within a niche.

### 5.3 Brief summary of results: what the case studies show

A review of these two examples, SunLine and CARTA, clearly demonstrates that in both cases niches contributed to the creation of the smallest possible parameter $N$. In Thousand Palms, this is achieved through the application niche of the bus market and the spatial niche of the SunLine transport company. In Chattanooga, the overall parameter $N$ is additionally delimited by the niche of the shuttle service. Thus, in both cases, the niches result in the new technologies impacting on the convention in an area that is very favourable for them.

The comparatively small bus fleet in Thousand Palms enabled the complete replacement of the vehicles in one fell swoop. In this way it was possible to overcome the predominance of a hyperselective convention and an early market for CNG buses was created. These had to prove themselves under market mechanisms and succeeded in doing so. Through the creation of new niches and extension of existing ones, the CNG buses were able to disseminate further and, since then, conquer an increasing proportion of the automobile innovation space.

The special features of the niche in Thousand Palms also created a solid basis from which they can expand to other areas. Thus, after the positive experiences of SunLine, the CNG buses were also introduced in Los Angeles and were gradually able to dominate the conventional technology there. It was planned to replace the remaining diesel buses in Los Angeles in 2008.

A multiple niche for an alternative technology with comparatively favourable initial conditions was also created in Chattanooga. Through the establishment of the niche of a shuttle service, the competition space was limited further. Thus the overall value of $N$ was further reduced. The battery buses only compete against the diesel buses under the specific conditions of the shuttle service.

The experiences in both Thousand Palms and in Chattanooga reflect the above-summarized results of the model formation in real-life conditions. In both cases, the competition space of the two technologies is significantly reduced through a double or multiple niche. Within the protection area of these niches, the dominance of the hyperselective convention is neutralized and the alternative technology has a realistic chance of becoming established.

## 6. Conclusion and future research

In this paper we have described the hyperselective attractor landscape of the automobile innovation space. We developed a matrix process with which this landscape is mapped. As part of this process, an innovation matrix is created that enables the systematization and classification of the innovations in this landscape [28]. An evaluation matrix can be derived from this innovation matrix which makes it possible to relate the different innovations and innovation types to each other on a comparative basis through parameters and to evaluate them [27]. While the innovation landscape extended along the two major technical dimensions (engine and fuel), it may also be said that it is a representation of the engineering knowledge space. The evaluation criteria fulfil aspects of effectiveness, marketability, usability and sustainability of propulsion innovation. As discussed by [57], barriers to the survival of new vehicles can emerge in the economics of the production process, at the level of required infrastructure or due to the infancy of technological solutions. Using the language of attractors and landscape, the problem of alternative propulsion could also be mapped to different landscapes: i.e. the space of technical solutions, the space of alternative fabrication, the space of competing infrastructures and the space of dominant social mobility pattern. This variety is maintained in the qualitative description of the situation and the classification scheme. However, for our innovation landscape we condensed the different aspects into dominant regimes which are identified by means of occurrence or occupation.

The visualization of the innovation landscape led us to the archetypical problem of how to switch between different relatively isolated islands in this landscape. In a further condensing step, we used a toy model of two competing technological regimes, which incorporates non-linear and stochastic aspects, for the discussion of possible escape strategies from a current lock-in in automobile technologies. In our view, stochastic approaches are indispensable to reaching an understanding of the process around the penetration and diffusion of a first singular event – i.e. a new technology. With regard to the birth situation of new technology models, in particular, models based on trend analysis can miss out important mechanisms governing how a technology can survive even in inopportune circumstances. We rely on a dynamic niche theory of the survival strategies of a new technology. In the final section of the paper we contrasted our theoretical scenarios with real-world observations using buses as a test case for pioneering new propulsion technologies. We presented arguments for fleet experiments which keep technological options open and provide a safe testbed for in-vivo testing on a small and medium-sized scale for alternative propulsion.

**6.1 Models and complexity**

Our use of models in this analysis differs in part from the usual role of mathematical models in the understanding and forecasting of technological innovations. In terms of the kinds of mathematical models to be used we can differentiate between models which try to be as realistic as possible and models which try to represent essential features of reality. It eventually boils down to a different approach to understanding complexity. In the one case the model is like an identical mirror of complexity. Complexity is transformed into the model world. Most of the system dynamic models, including climate change models, are of this kind.[2] The model enables the simulation of scenarios as the only form of experimentation possible for complex social systems. However, they also entail a lot of uncertainty. There are different possible sources for this uncertainty: the availability and accuracy of data to calibrate the model, the assumed underlying processes and mechanisms or the wide range and divergence of possible outcomes [58]. From an epistemic point of view, it is often impossible to gain insights into different principal scenarios, i.e. the model is as complex as reality.

---

2  For models on the future of energy and transport, see for example Energy Research and Investment Strategies (ERIS) [13] and Canadian Energy System Simulator (CanESS) [57].

The other extreme involves models which try to mirror principals or universal laws behind certain phenomena. They are as complex as required and as simple as possible [59]. A famous example here is the model of chemical oscillations (Brusselator) which is far from correct in relation to the actual reaction paths but is successful in terms of uncovering the primary driver behind oscillations, i.e. two substances involved a specific reactive feedback cycle. Other examples include the rule-based model of increasing returns developed by Brian Arthur [16] and Stuart Kaufmann's NK model [60]. These models are designed as thought experiments rather than real experiments. The model we used clearly belongs to this class. But this approach is also not without drawbacks. When designing models as selective mirrors or toy models the complexity of the real world is now transformed into the complexity of the selection of the primary elements and interactions. Complexity is now hidden in the choice for the representation. In our study we develop a niche concept close to a mathematical model. We gain operability through the identification of three major parameters for such a niche instead of having to deal with hundreds of possible indicators and influences. Empirical research is needed to validate, if we picked the right three parameters.

### 6.2 Models and data

In addition to the question as to how complexity should be dealt with, another issue that arises concerns the collection of data and the validation of models based on both quantitative measurement and qualitative observations. A model can be used in the analysis and explanation of a data-driven phenomenology or as a heuristic device for the ordering of observations and thoughts. In this paper we used our model in this heuristic manner. We measured where possible but, where exact data were unavailable, we also used arguments of plausibility or resonance with empirical observation. For example, we did not aim to map the innovation landscape in terms of indicators but based our arguments on a systematic derived from insight-based knowledge, interviews and document studies encompassing both qualitative and quantitative elements. It is our belief that there is a hidden rationale for our choice of model and its treatment in this analysis of alternative propulsion, and this concerns the usual audience for models.

### 6.3 Models in the interdisciplinary discourse

In the attempt to describe the hyperselective attractor landscape of the automobile innovation space not only metaphorically but with scientific accuracy in both qualitative and quantitative terms, we encountered two main directions in the current literature. These can best be described as "model-faithful" and "holistic" descriptions.

"Model faithful" descriptions move in the context of certain research traditions. They can build on statistical models, systems dynamics, game theory or – as in our case – the theory of non-linear dynamics. A different vocabulary is used in each case and expert knowledge is required to understand and evaluate the model statements. Despite the fact that it takes advantage of established insights, this kind of description has two crucial disadvantages: first, it is difficult, if not impossible, to unite different and often complementary models and model languages in one precise and uniform concept; second, the model concepts and, even more, their syntheses are more or less abstract. This can lead to the formation of communications blockades between the model experts, on the one hand, and the potential target groups of the modelling, on the other. Special translation work is usually required to convey the results of the analyses of other scientific disciplines to the political sphere and the general public.

"Holistic" descriptions are not focussed on specific models and the problem aspects behind them but on the most holistic view possible of the problem to be analysed. This includes both expert-cultural and everyday perspectives. Based on this, "holistic" descriptions are necessarily interdisciplinary in structure. They are positioned to a certain extent "across" the different "model-faithful" descriptions and try to connect the different analysis perspectives in such a way that neither scientific accuracy nor clarity and comprehensiveness suffer as a result. In our paper we developed such a holistic description and aimed to bridge the gap between the discourse around lock-in from a model perspective and the many detailed observations that exist in relation to lock-in situations.

It would appear that the use of models as "heuristic devices", as so-called toy models or thought experiments, is a necessary although insufficient condition for their integration into such a holistic description. This enables the co-evolution of models and observations in the framework of a holistic description. With our study we would like to advocate for model-inspired research which takes models and concepts from complexity theory neither only as a metaphorical point of departure nor as vehicles for a pure statistical data analysis. Instead we gave an example for a research which conducts and analyses qualitative and quantitative observations close to the mathematical apparatus of the model without aiming to obtain a one-to-one translation. We attempted to strike a balance between a high degree of accuracy on a conceptual level and a lower degree of accuracy on the level of data and measurement.

Technological change in the automobile industry is confronted with the fact that we need not only a product/end-product or process replacement, but a bundle comprising both and, moreover, it is needed from the very outset for any substitution infrastructure need to be taken into account. In this respect, propulsion innovation represents a classic case for network innovations. Last but not least: we know that technological change is socially shaped and the selection of one specific technological trajectory from the many technically possible options is ultimately a social choice [61]. The solvability of the engineering and technical problems at both local and network level is only one aspect. Moreover, economic factors are not the only crucial aspect: technological and market factors are not the only issues at stake in competition. What is involved is the mobilization of knowledge, information and decision networks at the most diverse levels of society. These are ultimately influenced and linked through the Z*eitgeist* and cultural streams [62]. This is clearly a complex phenomenon and we propose that complexity models should be used to obtain answers to the societal challenge of future mobility and to shape the political and societal conditions that foster the necessary change.

**Short bio of the authors**

Dr. Andrea Scharnhorst is a Senior Research Fellow at the Virtual Knowledge Studio for the Humanities and Social Sciences (VKS) at the Royal Netherlands Academy of Arts and Sciences (KNAW) in Amsterdam, Netherlands. Scharnhorst holds a Diploma in Physics and a PhD in Philosophy; her work is focussed at the interface between physics and social sciences/humanities. Since the 1990s, she has contributed to the growth of complexity theory in sociology and economics with models of growth and change of scientific fields, with developing paradigms for the survival of technological innovation and with learning agent models. She co-edited a volume with Andreas Pykla on Innovation Networks (2009) for Springer's Complexity Series .

Dr. Lutz Marz is a Research Fellow in the research unit, "Cultural Sources of Newness" at the Social Science Research Center Berlin (WZB) in Berlin, Germany. Marz' background in natural sciences and economics has been advantageous for his development of the *Leitbild*, a conceptual framework which is ideally suited to analysis of cultural and behavioural influences in organizations in the process of strategic decision making vis-à-vis technological innovations. Marz has published extensively in such fields as mobility research, innovations in future technologies – especially the automobile industry – future strategies of technological development and the changing role of China in technological innovation. Among his most important works is Visions of Technology: Social and Institutional Factors Shaping the Development of New Technologies (1996), co-authored with Meinolf Dierkes and Ute Hoffman.

Dr. Thomas Aigle is a manager at the Fuel Cell Education and Training Center Ulm (WBZU) in southern Germany. The WBZU provides high-level educational and vocational training in the field of fuel cells, batteries and hydrogen technology; its versatile program addresses different target groups ranging from staff in trade and industry, professionals in research and development, to students in higher education, and to the general public. Aigle's background in business and engineering has involved him in a number of important/groundbreaking projects in the field of fuel cells, alternative drives and alternative energy resources, on both the German national and the European level.